# Bayesian Comparison of Covid Origins Hypotheses

**Michael B. Weissman**

**Department of Physics, University of Illinois at Urbana-Champaign,**

**1110 W. Green St., Urbana, IL 61801, USA**

mbw@illinois.edu

Orcid 0000-0003-0081-4429

**Abstract**

Applying standard Bayesian methods to the most reliably known unclassified facts about the Covid epidemic allows crude estimation of the odds for research-related vs. purely zoonotic origins. Here a broadly inclusive zoonotic hypothesis is compared with a rather specific subset of the research-related hypothesis defined by known prior work and grant proposals. The important observations are the type of virus, the city of the first outbreak, the lack of known prior hosts, the anomalous early pattern of evolution, and two specific features of the viral sequence. The resulting odds strongly favor research. The odds remain fairly strong under inclusion of uncertainties in prior odds.

Acknowledgements: I thank Simon Wain-Hobson, Gilles Demaneuf, and Steve Massey for excellent editorial advice, but none are responsible for the final version. Others who have provided important advice wish to not be named.

## 1. INTRODUCTION

### 1.1 Purpose

Estimating the origins of the SARS CoV-2 virus (SC2) is the key to rational planning to avoid the risk of future pandemics. If the virus arose by some traditional zoonotic route involving wildlife and farming then it serves as a reminder of the need for monitoring and regulating practices in those fields. If it is the first example of a pandemic from a pathogen enhanced by research, then it serves as a forceful reminder that the growing field of potential pandemic pathogen research needs much tighter regulation. (A. Chan 2026)

### 1.2 Prior Work and Methods

#### *1.2.1 Modified Grunow-Finke*

(X. Chen et al. 2024) used a specialized method, a modified Grunow-Finke tool (mGFT) (X. Chen, Chughtai, and MacIntyre 2019) to compare a broad version of the lab-leak (LL) hypothesis with a broad version of the zoonotic (Z) hypothesis. They find that LL is favored. Although originally designed for a somewhat different comparison, deliberate release vs. Z, mGFT analysis has the virtue of using a mostly pre-specified set of features (X. Chen, Chughtai, and MacIntyre 2019) mGFT does not, however, have an intrinsic natural translation into an odds comparison, although it has been calibrated on a small number of cases.

#### *1.2.2 Bayes compared to mGFT*

The general-purpose systematic method of comparing the probabilities of competing hypotheses is Bayesian analysis (Gelman 2006) , constructed to produce odds estimates in an intrinsically logical way. Although it is in principle superior to the less quantitative mGFT tool, its lack of pre-specified types of observation to use can make it vulnerable to bias in selecting evidence. Thus a Bayesian analysis may be viewed as complementary to an mGFT analysis. Bayesian analysis cannot remove uncertainty or even fully remove subjective bias. Its great virtue is that the steps become transparent, allowing explicit disagreements and systematic modifications as new data or improved analysis become available.

### *1.2.3 Previous Bayes Analyses and Hierarchical Methods*

Several Bayesian analyses of SC2's origins have been more or less formally published. The nominally Bayesian Appendix to Steven Quay's book "The Code as Witness" (Quay 2026) contains a detailed discussion of many features that are unlikely for zoonosis but it lacks a systematic comparison to likelihoods for a lab leak. (Levin 2025) has posted a Bayesian analysis of Covid's geospatial properties, but it does not include priors or biological properties. A summary of a collection of varying Bayesian analyses appeared on Scott Alexander's blog. (Alexander 2024) Other analyses, some out-of-date, have been posted on other blogs. (Demaneuf and De Maistre 2020; Brunner 2021; Weissman 2025)

The odds obtained by these analyses vary widely, with most of the variance driven by a few extremely large likelihood factors. Extreme likelihood factors tend to show up when simplified models of outcomes from different hypotheses are used without taking into account the range of variants of each hypothesis, the possible unreliability of the data, and the uncertainties in the proper forms of the models by which the evidence is converted to likelihoods for the competing hypotheses and in the values of parameters used in the models. Hierarchical Bayesian analysis (Gelman 2006) can incorporate reasonable estimates of the reliability of each piece of evidence and of the uncertainty in the models. It is essential to take such uncertainty into account when comparing compound hypotheses using a broad range of different types of evidence of varying reliability for which there are no formal methods of deriving likelihoods. I will use informal hierarchical methods here, in effect down-weighting evidence for which a likelihood is hard to estimate. This acknowledgement of substantial uncertainty tends to reduce the odds ratios obtained from evidence. (In our calculation that will keep odds favoring LL over Z from becoming extreme.)

*1.2.4 Robust Prior Distribution*

Bayesian analysis requires some prior odds obtained by informal methods. A "robust" Bayesian approach (Berger, Insua, and Ruggeri 2012)should consider odds from a broad range of priors that reasonable people might hold. One way is simply to acknowledge that each evaluator will use somewhat different priors and thus get a different result. It is also helpful to give a sort of robust consensus result incorporating that uncertainty. A simple way to do that is to average probabilities obtained by integrating over a distribution of prior odds. That will pull our final odds toward 1.0. Averaging odds themselves over that distribution of priors would be incorrect and have the opposite effect, making the final odds more extreme.

There is no formal way to choose a good distribution of priors although some such choice is implicit in any evaluation, whether or not it is explicitly Bayesian. Here we shall see an uncertainty of around a factor of 10 either way in the prior odds. There is no well-defined model to guarantee that there are not bigger mis-estimates, so it is appropriate to use a fat-tailed distribution of ln(odds). I choose a 3-degree-of-freedom t-distribution with a standard deviation of ln(10). Figure 1 illustrates the effect of averaging over this distribution on the posterior ln(odds) as a function of the pre-averaged ln(odds), showing the reduced sensitivity to likelihood factors.

Figure 1. The posterior ln(odds) resulting after averaging over a 3-d.o.f t-distribution of prior ln(odds) with standard deviation ln(10) is shown. Plotted with help from Gemini.

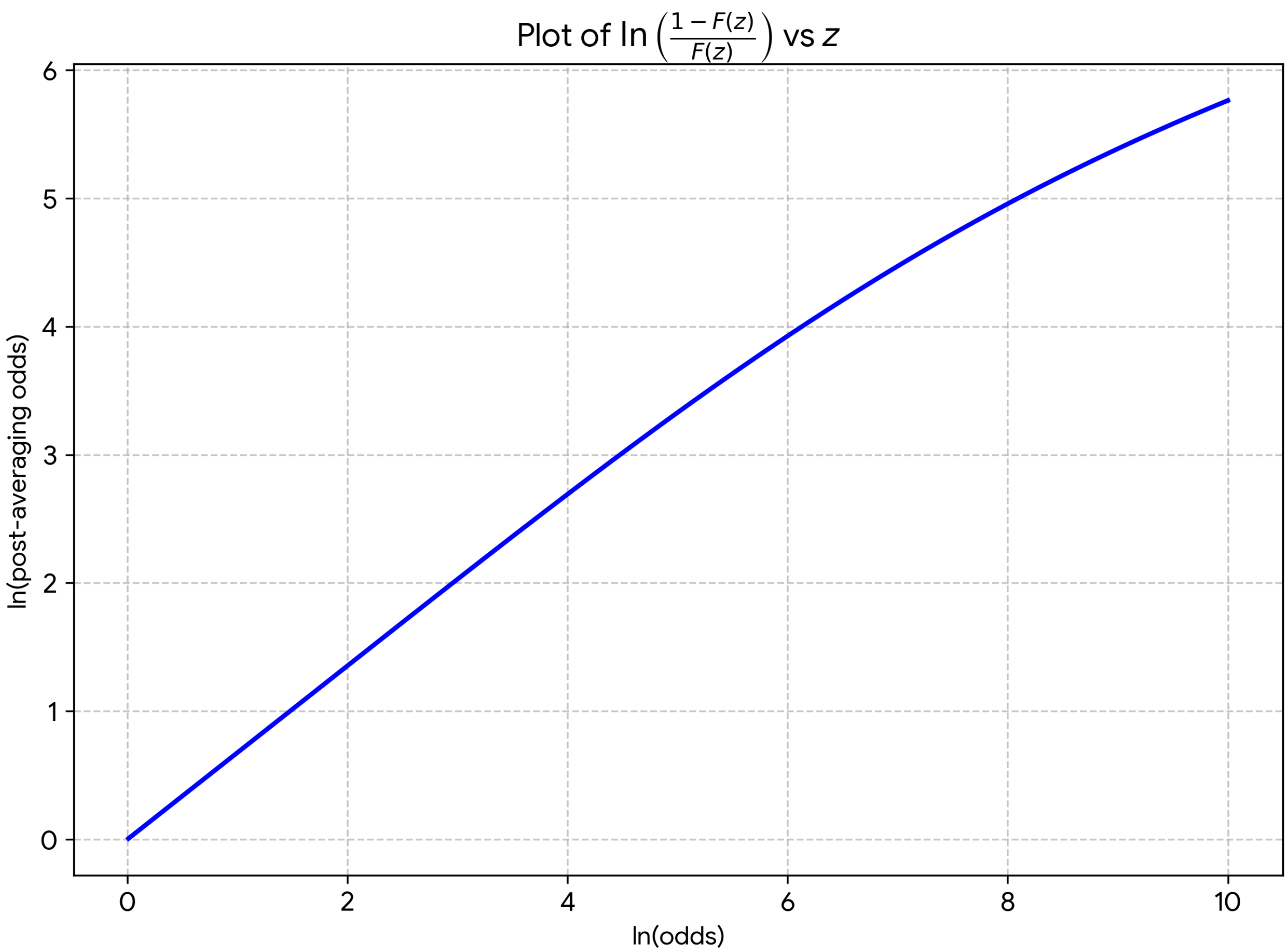

*1.2.5 Focusing on a Sub-Hypothesis*

The analysis here will be greatly simplified by restricting the LL hypothesis to one subset, research along the lines outlined in a 2018 grant proposal, "DEFUSE" (Daszak 2018) submitted to the Defense Advanced Research Project Agency by the EcoHealth Alliance, the University of North Carolina (UNC), the Wuhan Institute of Virology (WIV), and the Duke National University of Singapore. DEFUSE proposed studying sarbecoviruses, viruses fairly closely related to SARS CoV-1 (SC1), by creating a series of chimeras of related viruses and seeing which genetic modifications affected the growth in cell cultures and then in humanized mice. Those mice have a genetic modification making the protein (ACE2) by which the virus binds to cells similar to human ACE2.

Limiting to a fairly specific version of LL allows relatively well-defined priors and likelihood factors compared to fuzzier estimates involving many labs whose activities and plans are nearly unknown and ultimately irrelevant. Although dropping any chance for LL to arise from dissimilar projects and distant labs in principle reduces the LL odds that will be unimportant here.

*1.2.6 Core Data for Bayes*

I will confine the analysis to the most obvious "what, when, where" features that any analysis would use, plus features for which at least the direction of the likelihood ratio unambiguously points one way or the other. I will avoid complicated modeling, in which implicit assumptions tend to be hidden. The resulting factors can be viewed as a snapshot of the most reliable current information, suitable as a starting point for future additions, subtractions, and modifications.

(Bannan 2026) outlines the main evidence on the origins question. Our outline is close to his, but fleshed out with a rough Bayesian odds estimate and including one other sequence feature.

Every conditional probability here will include "giving rise to a pandemic" as an implicit part of each hypothesis. That matters because the subsets of hypothetical zoonosis or lab leak accounts that give rise to pandemics are different in relevant ways from more inclusive sets. This implicit conditioning is particularly important to avoid underestimating the conditional probability under Z of features that contribute to the viral fitness.

## 2. PRIOR ODDS

### 2.1 The Line Between Priors and Likelihoods.

The line between features to include in the prior probabilities $P_0(Z)$ and $P_0(LL)$ or in the likelihood factors can be chosen for convenience. Here I use the probability of a successfully propagating new sarbecovirus disease of humans for the prior probabilities because it avoids using unnecessarily fuzzy factors for $P_0(Z)$. Some factors going into $P_0(LL)$ and $P_0(Z)$ that might in other analyses be included as likelihood factors, e.g. P(sarbecovirus|LL)/P(sarbecovirus|Z), will be included here in $P_0(LL)/P_0(Z)$. Other factors that some might include in the priors, e.g. P(Wuhan_location|LL)/P(Wuhan_location|Z) are separated out here as likelihood factors. Each $P_0$ here will implicitly be limited to a one-year period, 2019.

### 2.2 Prior for Zoonosis

$P_0(Z)$ can be crudely estimated from a simple fact, that there has been one known zoonotic sarbecovirus disease in humans, the 2003 SARS epidemic, SC1. As always, the event in question, SC2 itself, cannot be used as part of the prior training set since making any assumption about its origin would amount to circular reasoning. There is no special reason to assign strong time-dependence to the probability of such disease from zoonotic sources. One cannot, however, reliably extend the assumption of roughly time-independent risk back to periods with much lower population in the area where sarbecoviruses flourish, Southeast Asia. We can conservatively approximate:

$P_0(Z)$ = ~1 year/20 years = 1/20.

This estimate would be lowered if a narrower sarbecovirus category were used. DEFUSE specified a virus with spike protein 10%-25% different from SC1. UNC's Ralph Baric testified to Congress that the research group's interest focused on strains whose percent difference from SC1 was "a number between 20 and 25. And that was our prediction, that there would be strains with that much variation that could still use human ACE2 receptors. It turns out SARS1 had 22% variation." (Committee on Oversight and Responsibility 2024)

**2.3 Prior for Lab Leak**

$P_0$(LL) will be even more crudely estimated by a more complicated argument. One part is simple: the DEFUSE plan was devoted to experiments on sarbecoviruses, so any leak from DEFUSE-related research be of a sarbecovirus. To estimate the yearly probability of such a leak we can turn to prior estimates of research risks. In calculating $P_0$(LL) one needs to take the product of the probabilities that one of the research groups involved pursued that line of research, that it would produce a highly transmissible virus, and that the virus would leak from the lab.

The probability that research along DEFUSE-lines would continue is not much less than 1. The WIV branch of the DEFUSE team continued to receive closely related funding from the NIH (Daszak 2018) and after DEFUSE was turned down received a grant on about the same topic from the Chinese Academy of Sciences. (Chinese_Academy_of_Sciences 2018) According to CIA intercepts, pangolin coronaviruses were sent from Beijing to Wuhan for use in experiments in transgenic mice. (Fauci 2026)An unpublished sequence for an artificial MERS virus has showed up as a contaminant in agricultural samples from Wuhan (Jones et al. 2024) confirming that work on unpublished sequences was ongoing.   When Der Spiegel asked Shi Zhengli, the lead coronavirus researcher at WIV, whether the work had started anyway she responded “I don’t want to answer this question…” (Bratlie 2025) We may take the probability of a project like this going ahead to be ~0.8.

The probability that a highly transmissible virus would leak in a one-year period is also not very low. A series of warnings of the likelihood of leaks preceded the pandemic. (Klotz and Sylvester 2012; Casagrande 2015; Butler 2015; Lipsitch 2018; Johns_Hopkins_Center_for_Health_Security 2019; Feng et al. 2019)

A typical prior estimate of the probability based on records of previous leaks is ~1%/year per full-time lab worker, assuming the work is carried out at BSL-3. (Lipsitch and Inglesby 2014) Based on the scale of the grants there would have to have been a small team assigned to this project.

Exchanges between DEFUSE participants (Kopp 2023)showed that they planned to use "cost effective" measures by conducting much of the work in Wuhan at insecure BSL-2 rather than the more secure BSL-3, although Baric warned “In China, might be growin these virus under bsl2. US researchers will likely freak out.” [sic]. Alarms about safety issues at Wuhan labs had already been raised. (Rogin 2020) SC1 repeatedly leaked from labs despite a requirement of BSL-3 safety, e.g. from a lab in Singapore (Lim et al. 2004) at least twice from a highly prestigious Beijing lab (Walgate 2004), and even from a BSL-4 lab in Taiwan. (Orellana 2004) That led WHO to conclude that for SC1 "The risk of re-emergence from a laboratory source is thought to be potentially greater", compared to that "from an animal reservoir" (WHO 2006). Unlike SC2, SC1 had little or no asymptomatic transmission, making it easier to prevent SC1 leaks from spreading. The probability that the highly transmissible SC2 would then leak from a DEFUSE-like project at BSL-2 is likely at least 0.1/yr.

The biggest uncertainty in crudely estimating $P_0(LL)$ comes from estimating the probability that a lab-modified virus would not only be human-transmissible but also that it so transmissible would successfully spread, creating a major problem.  Although for evaluating general risks one should include less transmissible pathogens that might acquire more transmissibility in circulation, for evaluating the likelihood of LL for SC2 specifically  we would restrict that to viruses that quickly started rapid spread, Research on lab-modified viruses has only ramped up in recent years, so there is no good historical record to use to estimate this probability. Previous work from a group including two DEFUSE leaders (UNC's Baric and WIV's Shi) had created chimeric sarbecoviruses with replication in human cells and humanized mice enhanced enough that the authors acknowledged "scientific review panels may deem similar studies building chimeric viruses based on circulating strains too risky to pursue, as increased pathogenicity in mammalian models cannot be excluded." (Menachery et al. 2015)The key feature enhancing the human infectivity of SC2 compared to previous sarbecoviruses is its furin cleavage site (FCS), but since DEFUSE proposed adding such sites the probability under LL of having one is high.

The path that the DEFUSE researchers believed might succeed in making a dangerous pathogen was outlined in a report describing Baric's talk at a 2019 workshop, pages 10-11 of (Harbin_Veterinary_Research_Institute 2019). "Studies to alter pathogen properties of viruses can use several approaches including selection pressure...as well as deliberate design.... But he cautioned that a combination of techniques, including selection pressure from passaging in mice, is generally needed to generate virulent strains in a host and reminded the audience of the law of unintended consequences and the possibility of surprise." Baric also pointed out that maximizing virus-host molecular interactions was often not the path to maximal viral fitness, a point that is relevant to some claims (Andersen et al. 2020) that SC2 could not have been modified in a lab. We may crudely estimate the probability of success in making an especially transmissible chimera as 0.1. The uncertainty in this estimate is the main contributor to the uncertainty of prior odds.

### 2.4 Prior Odds

Combining these estimates gives:

$P_0(LL) = \sim 1/140$

That gives prior odds

$P_0(LL)/P_0(Z)$ = ~1/7.

These odds seem less favorable to Z than one might intuitively expect solely because they already are limited to a narrow type of pathogen, sarbecoviruses. Inclusion of all types of new pathogens would raise $P_0(Z)$ by about an order of magnitude, since many potential zoonotic examples are found even in the narrow category of viruses in farmed fur animals in China. (Zhao et al. 2024)

**3. WHERE**

**3.1 Likelihood for Lab Leak**

The first reliably detected outbreak was in Wuhan China, the location of the first officially identified cases, the first distinct increases in hospitalization and death, and the basal lineages of all subsequent sequences. Since we have already restricted the LL hypothesis to the narrow category of DEFUSE-related work $P_0(LL)$ already paid the price of that restriction so P(Wuhan|LL) is high. The risky part of that work would have been conducted either in Wuhan or at UNC, with most of the risk concentrated at the relatively unsafe Wuhan labs. The first outbreak of a leak from a source in a large city would almost certainly be in that city. Conservatively,

P(Wuhan|LL) = ~0.8.

**3.2 Likelihood for Zoonosis**

For P(Wuhan|Z) there is a much larger range of possible starting locations. Although for Z the presence of surveillance at Wuhan might tend to make it more likely than a random spot for the first detected cases, that would not account for it being the location of the first big outbreak which could have been observed in any urban center. Wuhan's population is well under 1% of the population of the East Asian countries in which sarbecoviruses are found and roughly 1% of

the urban population.  Wuhan is not an especially likely place for a new sarbecovirus to emerge since it is over 1000km from the region in which the most relevant viruses are found, southern Yunnan, Laos, and Vietnam. (Hassanin et al. 2024; Jonathan E Pekar et al. 2025) Sarbecoviruses in general have a broader, multi-continent distribution. WIV had even previously used Wuhan residents as a negative control in serology studies of exposure to sarbecoviruses.  (N. Wang et al. 2018) Shi Zhengli initially reacted "I had never expected this kind of thing to happen in Wuhan, in central China." (Qiu 2020)

We can then estimate

P(Wuhan|Z) = 1/100.

**3.3 Likelihood Ratio**

This gives a likelihood factor

P(Wuhan|LL)/P(Wuhan|Z) =80.

One widely publicized version of Z, denoted $Z_{HSM}$ here, involves the wildlife trade at the Huanan Seafood Market (HSM). Wuhan's share (Xiao et al. 2021) of China's wildlife trade (Chinese_Academy_of_Engineering 2017) is an order of magnitude or more lower than its share of China's population, so a correspondingly lower P(Wuhan|$Z_{HSM}$) would be needed for that version.

**3.4 Interim result**

Just using that SC2 was a sarbecovirus that first showed up in Wuhan we have obtained odds favoring LL by ~80/7. After averaging over the distribution of priors, we find ~5/1, not far from the corresponding base estimate of (Demaneuf and De Maistre 2020).

## 4. NO HOSTS FOUND

### 4.1 The Data

No wildlife intermediary has been found anywhere for SC2 despite intense searches. According to the Lancet Commission (Sachs et al. 2022), relying on data from the Chinese government, "Despite the testing of more than 80000 samples from a range of wild and farm animal species in China collected between 2015 and March, 2020, no cases of SARS-CoV-2 infection have been identified." No animal in the general vicinity of Wuhan was found to carry any close relative of SC2. (Wen Wang et al. 2022)

### 4.2 Likelihoods

For LL one would generally expect to find no external prior animal host since, as we shall see, the most likely versions of LL involve some evolution or modification of the virus in the lab. If the virus happened to be a nearly unmodified one obtained from sampling remote locations it would be unlikely that the lab would choose to reveal that source. Thus P(no_host|LL) = ~1.

We can roughly estimate P(no_host|Z) by looking at other coronaviruses. For both the important recently spilled-over human coronaviruses, SC1 and MERS, intermediate wildlife hosts were found, as were hosts for 3 of the 4 other recently identified human betacoronaviruses. (Schindell et al. 2022) It would not be likely that anyone could have found the intermediate host for the missing one, HCoV-HKU1, which causes a relatively minor disease (X. Chen et al. 2022) since retrospective evidence indicates its existence in Brazil and elsewhere (Góes et al. 2011) years before it turned up in Hong Kong, with no evidence of search for hosts there. A broader review of human coronaviruses finds that probable intermediate hosts have been identified for 5 of the 7 described, not counting SC2.(Tang, Liu, and Chen 2022) HCoV-NL63 is the other human coronavirus whose probable proximal host has not been identified, but since "HCoV-NL63 has been present in the human population for centuries" (Abdul-Rasool and Fielding 2010) and produces mild disease there has been little opportunity or motivation to

search for its prior host.

Given the enormous attention paid to SC2, the known location of its initial outbreak, and the improved sensitivity of detection, P(no_host|Z) would be less than for the other coronaviruses. With 5 identified hosts and only two missing, both in cases where there was no search, a conservative estimate would be

P(no host|LL)/P(no host|Z) > 7/2.

4.3 **No Multiple spillovers**

As (Bannan 2026) points out, the single location of the known spillover of SC2 contrasts with the repeated multiple spillovers of both SC1 and MERS. Multiple spillovers would make it easier to track a prior host and pre-adaptation would make it easier for a pandemic to start without multiple spillovers, so to avoid double counting I will not use the lack of multiple spillovers to obtain an additional likelihood ratio.

4.**4 Interim result**

Before considering any detailed information about the RNA sequence and evolution, we have found LL favoring odds of ~40/1. Averaging over our distribution of prior odds would reduce that to ~10/1.

**5. PRE-ADAPTATION AND INITIAL EVOLUTION**

**5.1 Likelihood for Zoonosis**

The peculiarly good initial adaptation of SC2 to human cells was immediately noted as atypical for zoonotic introductions, in particular by some of the authors of the "Proximal Origin" paper (Andersen et al. 2020) that was important in establishing the view that a research origin was unlikely. Lead author Kristian Andersen noted on 2/08/2020: "what we're[sic] observed is completely unprecedented as far as I know. Never before has a zoonotic virus jumped into humans and spread through the population like wildfire with the[sic] kind of speed." (Rambaut et al. 2020) Coauthor Eddie Holmes remarked on 2/10/2020 "It is indeed striking that this virus is so closely related to SARS yet is behaving so differently. Seems to have been pre-adapted for human spread since the get go." (Rambaut et al. 2020) To the extent that these virologists' subjective first impressions of pre-adaptation atypical for zoonosis are supported by the subsequent evolution of and analysis of SC2, that pre-adaptation would be less consistent with spillover from a wild species than with prior culture in human cells or humanized lab mice.

Those subjective impressions have held up fairly well. The initial protein evolution in humans (Markov et al. 2023) was much slower than for SC1 (Song et al. 2005) , with about a factor of 5 lower ratio of non-synonymous to synonymous mutations. The FCS region of the original SC2 also evolved little when grown in human cell cultures. (Lamers et al. 2021) These features suggest that the early stage of the pandemic was not the first opportunity SC2 had to adapt to human cells.

(Porter et al. 2023) checked whether the slow early evolution in humans was due to some special generalized cross-species infectivity of SC2 by comparing with early evolution in minks after spillover from humans. They concluded "[SC2's] apparent neutral evolution during the early pandemic….contrasts with the preceding SARS-CoV epidemics….Strong positive selection in the mink SARS-CoV-2 implies that the virus may not be pre-adapted to a wide range of hosts." A similar result was reported by (Tai et al. 2022) Initial evolution in laboratory cats, dogs, hamsters, and a ferret was faster than initial evolution in identified human cases. (Bashor et al. 2021) So was evolution after spillovers to tigers, lions, and hyenas in zoos (Bashor et al. 2025) and spillover to wild white-tailed deer. (McBride et al. 2023)

One possible route to pre-adaptation in the ascertained cases would be via undetected prior rural circulation of some initially less human-adapted version. (Bannan 2026) has pointed out that China's nationwide surveillance turned up "No cases of SARS...other than the 2012 reports of six cases of SARS illnesses in miners working in Mojiang county, Yunnan." Samples from those cases were transported to WIV where they provided the topic of a 2013 thesis. (Rahalkar and Bahulikar 2020)Surveillance reported no related illnesses in the years 2013-2018. (Bannan 2026) Retrospective serology studies of samples taken in late 2019, including almost 10,000 from patients with influenza-like illness, described on page 26 of (WHO-China_study 2021), found uniformly negative results. Samples from the early days of the pandemic found about two orders of magnitude higher positivity rate in Wuhan than either in Shijiazhuang to the north and Shenzhen, to the south and closer to the region with natural relatives. (Chang et al. 2021). Thus prior circulation in humans before reaching Wuhan is unlikely.

**5.2 Likelihood for Research**

SC2's atypical pre-adaptation should shift the odds toward LL since DEFUSE (Daszak 2018) described planned passaging in human airway epithelial cells and humanized mice.

**5.3 Rough Likelihood Ratio**

Although the early evolution in humans stands out as anomalous against a background of 9 other species spillovers, to be conservative with this factor let us say P(pre-adaptation|LL)/P(pre-adaptation|Z) > 3.

5.5 **Interim result**

Including the anomalous early evolution, we have found LL favoring odds of ~120/1. Averaging over our distribution of prior odds would reduce that to ~25/1.

**6. SEQUENCE FEATURES**

**6.1 Features and Fitness Constraints**

Two features of the SC2 genetic sequence are surprising for zoonotic origins but not for DEFUSE-style research products. The most noted one is the FCS, not found in any other sarbecovirus. The FCS is important for the high transmissibility and pathogenicity of SC2 (Morgan et al. 2025; Peacock et al. 2021) but that makes it tricky to use its presence to calculate a likelihood because our observation itself is conditioned on there being a pandemic. I will use some peculiarities of the FCS to calculate a modest likelihood ratio. The other sequence feature is a special pattern of restriction enzyme (RE) cutting sites, less familiar but with the virtue of being irrelevant to fitness. The likelihood estimates for each of these features require more judgment calls than for the preceding ones.

### 6.2 The FCS

Many proponents of the LL hypotheses have claimed that the presence of the FCS is very strong evidence for LL since no other sarbecovirus has an FCS and SC2 has one exactly at the spike protein's S1/S2 boundary, where DEFUSE drafts discussed introducing one, as illustrated in Figure 2. Discussions among virologists who then became public advocates of the Z hypothesis show that they had the same first impression. For example, even before the DEFUSE drafts were released, one Proximal Origin coauthor, Robert Garry, initially reacted: " I really can't think of a plausible natural scenario where you get from the bat virus or one very similar to it to where you insert exactly 4 amino acids 12 nucleotide that all have to be added at the exact same time to gain this function -- that and you don't change any other amino acid in S2? I just can't figure out how this gets accomplished in nature.... Of course in the lab it would be easy to generate the perfect 12 base insert that you wanted".
(Committee_on_Oversight_and_Government_Reform 2023) Some care is needed to convert that widespread initial reaction to a rough likelihood ratio because the entire discussion is conditioned on the existence of a pandemic and the FCS is a contributor to that condition.

Figure 2. An illustration from the DEFUSE proposal of the research interest in effects of an FCS at the S1/S2 boundary. The other protease sites listed are ones already present in sarbecoviruses.

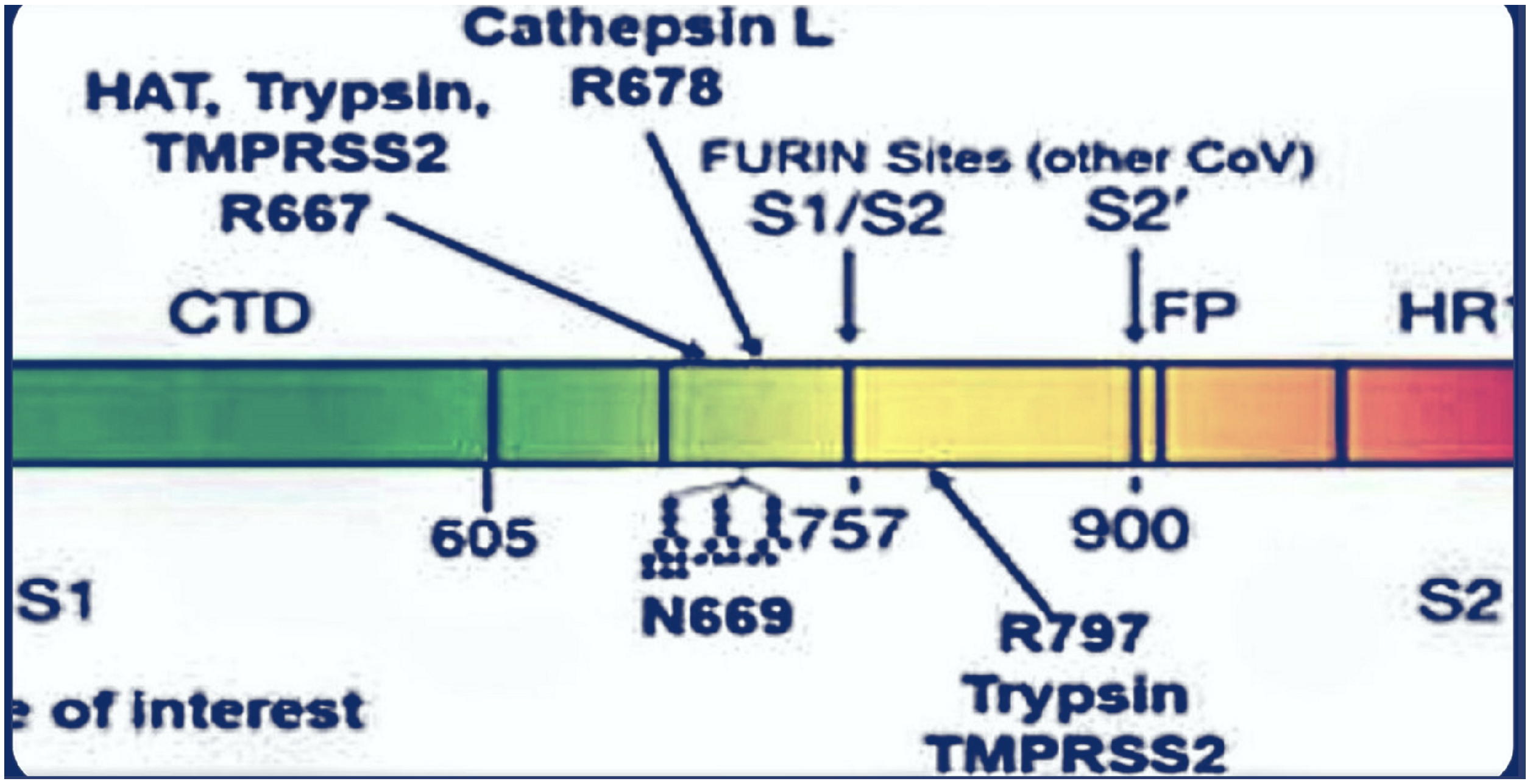

Garry's impression that the FCS arose via a 12 nt insert has held up well, but not the assumption that the insert could not have been natural. The particular coding CGCCGG for the arginine pair RR in the FCS is extremely rare in coronaviruses. (Romeu 2021) The only natural route to a reasonable probability for finding such coding would be in a recent long insert (Garushyants, Rogozin, and Koonin 2021; Bianco et al. 2026), for which the fraction of R's coded CGG is a surprisingly high 5.3% in samples from humans. (Hisner 2026) Sharper evidence of the recency of the FCS insert is provided by the rapid rise a few months into the pandemic of the D614G mutation that is strongly favored once the FCS is present but was absent in the initial human sequences. (Guo et al. 2022) Thus the recent insert path would be shared by plausible Z or LL accounts of the FCS.

Addition of an FCS at S1/S2 is not rare in virology research. (Belouzard, Chu, and Whittaker

2009; Follis, York, and Nunberg 2006; Y.A. Chan and Zhan 2022) A long-time collaborator of WIV used a 12 nt insert in a plasmid to make one. (Zhou et al. 2013) In Oct. 2019, Vincent Munster of NIH/NIAD proposed that it "[might be cool" to add a "broader cleavage signature (like furin-like) to enhance infectivity". (Munster 2019) His correspondent replied "Seems like more groups will be converging on this idea soon." (Munster 2019) After seeing the SC2 sequence, on Feb 1, 2020, Andersen agreed with his colleagues Holmes and Garry, "Yeah, the furin site would be the first thing to add for sure." (Rambaut et al. 2020) With DEFUSE drafts calling attention to the specific possibility of studying the effects of adding a "human specific" FCS at S1/S2, as shown in Fig.1, P(recent insert FCS|LL) would not be very low, particularly after conditioning on the existence of the pandemic.

The particular amino acid sequence of SC2's FCS, RRAR, matches the FCS of the familiar widely present human ENaC enzyme and thus fits the "human specific" specification. (Anand et al. 2020; Harrison and Sachs 2022). On the other flank, the inserted proline matches one on the human MERS virus that was being studied at WIV, (J. Chen et al. 2023) which also has an extended loop at the S1/S2 boundary, like that created by the 12 nt SC2 insert.(Bashir et al. 2024) Thus it is unlikely that DEFUSE-like research would have made a pandemic virus by some route other than an FCS. A lab could have made an FCS by point mutations rather than by an insert, although it is not clear that such an FCS would be functional.
P(recent-insert FCS|LL) = ~0.5 would then be a conservative estimate.

Since whatever predecessors existed in either nature or a lab would have gone through an FCS-favoring selection on the way to becoming a widely discussed pandemic pathogen, the prior FCS prevalence in bats cannot be directly used to estimate P(FCS|Z). The lack of an FCS in any non-bat sarbecoviruses does provide some evidence. Although the FCS of SC2 clearly has provided major evolutionary advantages for transmission in other species, the long period of bat interactions with a range of other non-bat mammals has not produced a spillover of a persistent FCS-containing sarbecovirus even though it has produced several successful spillovers (i.e. ones leading to detectable ongoing transmission chains) to non-bats, specifically two pangolin species, raccoon dogs, palm civets, and incidental spillovers to wild boar and ferret badgers. (B. Liu et al. 2023) Likewise the initial human SC1 did not pick up an FCS in its thousands of cases or in its spillovers to civets. (Cherry and Krogstad 2004) Each of these non-bat viruses have had chances to pick up an FCS both during initial spillover and during subsequent propagation. The first detected SC2 in humans already had an FCS despite having more limited chances to pick one up. A simple estimate of P(FCS|successful spillover) can be obtained by assuming a uniform prior on it and integrating over the likelihood that 0/4 distinct known spills would produce an FCS. That gives P(FCS|Z)= 1/6.

In addition to being able to spill over to non-bat species without acquiring an FCS, natural viruses have other routes to acquiring an FCS rather than just via a recent large insert, e.g. via gradual accumulation of small mutations. Thus P(recent insert FCS|FCS,Z) < 1. Overall then, we may consider P(recent insert FCS|Z) =1/6 to be a conservative estimate so we can conservatively estimate that the FCS gives
P(recent insert FCS|LL)/P(recent insert FCS|Z) > (1/2)/(1/6) = 3.

Further details would tend to further favor LL over Z as an account for the FCS, although not reliably enough to derive a Bayes factor here. A natural FCS, unlike the plan described in DEFUSE, would probably not be of a familiar human version. It would in general not be flanked by a proline, unlike the MERS virus under study at WIV. (J. Chen et al. 2021) (Lisewski 2024) has pointed out further SC2 sequence details previously found only in a lab-created MERS virus, but it is unclear if a lab would be doing that detailed engineering. In at least some synthetic sequences, the mRNA vaccines, CGG is by far the most common arginine codon (Xia 2021), although it is not especially rare in large natural inserts. On the other hand, assuming there was no mutation from related sequences before the insert, the insert was out-of-frame, which would happen ~2/3 of the time in nature but seems eccentric for a lab insert (Committee on Oversight and Responsibility 2024), although not increasing technical difficulty and perhaps having some engineering utility. (Lisewski 2024; Segreto and Deigin 2020)

Another sequence feature, an ablated glycan due to a point mutation T372A, appears to also be essential or nearly essential for easy human transmission. (Kang et al. 2021) DEFUSE does discuss modifying glycan sites but does not focus specifically on that one. Since that mutation is important for fitness but might potentially have been replaced with other glycan changes either in nature or in the lab, no large Bayes factor can be obtained from it. The apparent need for both the T372A and FCS mutations may reduce the probability of a successful natural spillover, but that effect would largely already be accounted for in the low $P_0(Z)$.

6.3 **Interim summary**

At this point we have conservatively approximated the Bayes factors corresponding to the basic points of (Bannan 2026), obtaining net odds ~400/1 favoring LL, giving ~50/1 after averaging over uncertain priors.

**6.4. The restriction enzyme site pattern**

(Bruttel, Washburne, and VanDongen 2023) claimed that SC2 has a pattern of cutting sites for the restriction enzyme pair BsmBI/BsaI that would be characteristic of synthetically assembled coronaviruses. The National Center for Medical Intelligence had independently noted the same pattern. (National_Center_for_Medical_Intelligence 2020) We shall see the probability of getting the specific pattern in nature is very low. Natural selection has no systematic effect on that probability. Laboratory design along the lines of the DEFUSE plans offers different pathways with higher probability of getting the pattern, as illustrated in Figure 3.

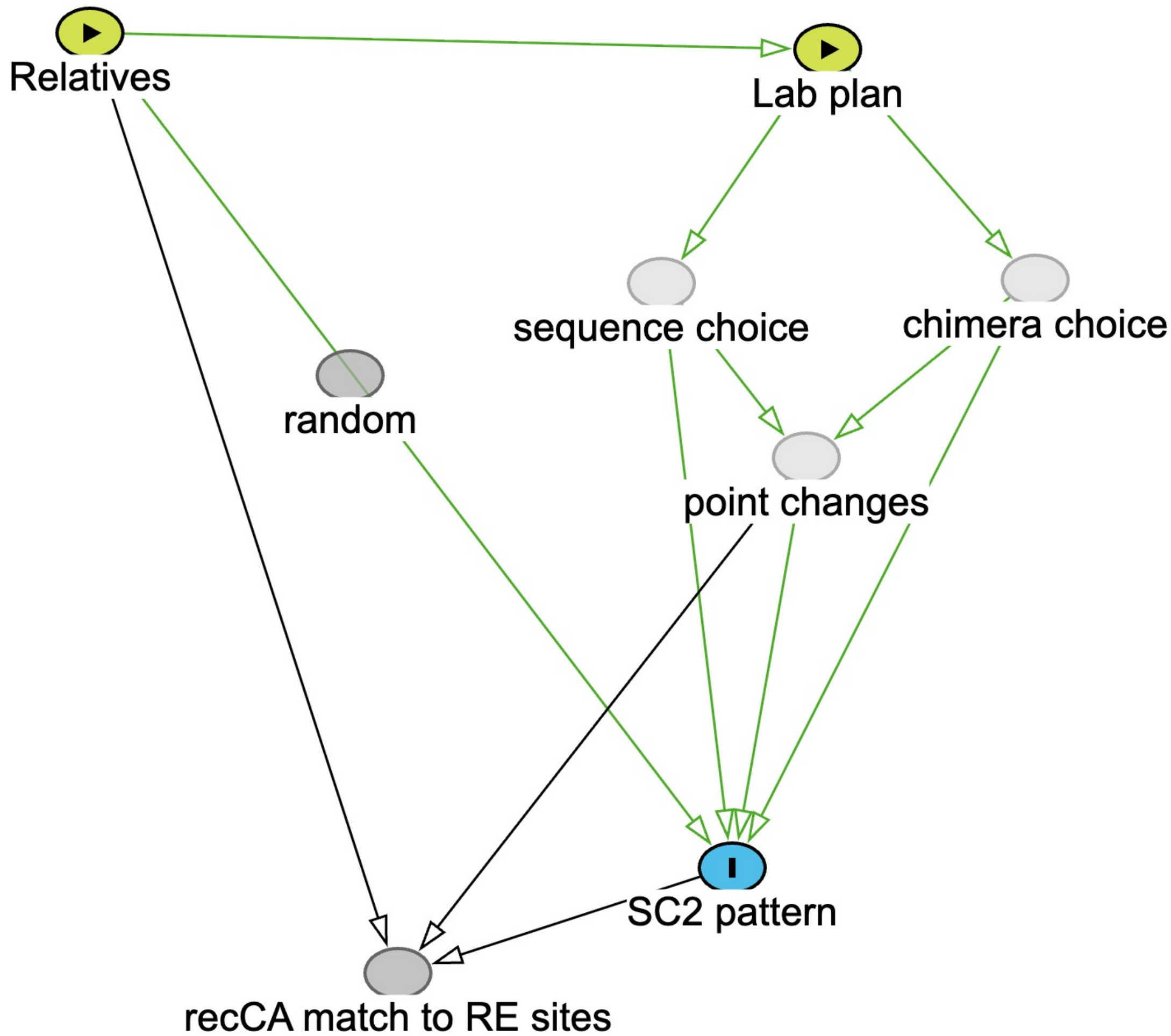

Figure

Figure 3 sketches the causal paths by which the RE pattern of SC2 might have arisen and of how its estimated recCA could have been determined. Directly observed properties are colored and theoretical constructs are gray. Only a fraction of the probable relatives have been observed. "Point changes" refers to possible lab changes in the potential RE sites to create the pattern. The match between RE site neighborhoods in SC2 and the probable estimates of the recCA is the same for all paths that don't go through LL "point changes". The direct LL path through "sequence choice" will always give likelihood greater than or equal to the likelihood for Z.

Since many restriction enzymes might be used and there are several other free parameters some knowledge of the research plans was needed to define the pattern well enough to estimate P(pattern|LL). The picture clarified when draft notes for the DEFUSE project were found (Kopp 2023) to propose to

> ... synthesize the genome using commercial vendors (e.g., BioBasic, etc.), as six contiguous cDNA pieces linked by unique restriction endonuclease sites that do not disturb the coding sequence, but allow for full length genome assembly....Using the full length genomes, we will re-evaluate virus growth in primary human airway epithelial cells at low and high multiplicity of infections and in vivo in hACE2 transgenic mice, testing whether backbone genome sequence alters full length SARSr-CoV pre-epidemic or pathogenic potential in models of human infection.

Although the RE sites are often removed in the final synthetic product, Bruttel et al. hypothesized that they might have been left in to facilitate ongoing creation of new chimeras by swapping in new segments since the DEFUSE proposal says “We anticipate recovering 3-5 full-length genome viruses/yr.” The largest DNA segment size conveniently available from commercial sequence vendors is 8 kb, so segments larger than that would not fit the pattern The plan to choose "sites that do not disturb the coding sequence " indicates that the sites were to be left in, since removed sites cannot affect coding. The RE supplier mentioned in the DEFUSE budget, (New_England_Biolabs) says of this RE category "The advantage of using Type IIS enzymes for assembly is that the recognition sequence can be... retained in the construct and can subsequently be reused”, as would be useful in making DEFUSE's "3-5 full-length genome viruses/yr". Re-use is described in, e.g., (Weber et al. 2011).

The DEFUSE budget notes (Kopp 2023)specify only one RE, a "Type IIS restriction endonuclease that ..replaces BsmBI". (New_England_Biolabs) Although no other RE is specified in the budget, swapping new pieces in for new chimeras could be facilitated by allowing separately swapping different parts by using a second RE, rather than requiring all the cuts to be made and segments reassembled. BsaI, another Type IIS enzyme, was already in use along with BsmBI at WIV for sarbecovirus research. (Ben Hu 2017) A subsequent WIV patent (Shi Zhengli 2023)describes breaking up a MERS sequence into segments using two BsaI sites together with seven BsmBI sites, reminiscent of the two BsaI sites and three BsmBI sites of SC2. The MERS clone that turned up in an agricultural sample had an anomalously low total number of BsaI and BsmBi sites, 2, indicating that in that case such sites were removed rather than left in for reuse.

Under LL, the observed pattern of 6 segments each less than 8 kb, (6, <8k), would arise so long as the researchers followed the plan to use modules with the RE sites left in for reuse, stuck with the planned segment number 6, and chose their customary BsmBI/BsaI enzyme pair. P(pattern|LL) then will not be very low, since the observed pattern closely fits the plans. We do not, however, know whether under LL the sites would have been left in, as the plan seems to indicate and is used in one type of assembly (Almazán et al. 2014) sometimes used by WIV (Zeng et al. 2016) or removed, as in the MERS clone. (Jones et al. 2024)Since we would not flatly reject the LL hypothesis if no such pattern were found or if it deviated slightly (e.g. with 7 segments), it would not be consistent to assign a probability of 1 to P(pattern|LL) since the sum of the probabilities for the pattern and variants cannot exceed 1.

Under Z there is no reason to expect to find any such plan, much less one that exhibits such a detailed fit with the observation. No one supporting the Z hypothesis indicated that they thought such a plan was plausible prior to its publication. A recent zoonotic-favoring analysis of the pattern (Hensel 2025)still dismisses the relevance of the DEFUSE plan and does not mention the WIV patent. It is, however, hard to evaluate just how low P(plan|Z) would be, although it is surely much lower than P(plan|LL), where "plan" here includes the DEFUSE reusable 6-segment plan including BsmB1 together with the WIV patent combining BsmB1 and Bsa1.

The large number of sequenced real viruses allows an empirical estimate of P((6,<8k) (Quay 2026) points out that the simplest feature of the pattern, the small total number of BsaI and BsmBI sites, is already rare regardless of their spacing. A pseudonymous analyst (Gadboit 2026) finds that only 2 of the 530 related viruses for which he has sequences have 5 sites with another 2 having fewer, indicating P((6,any lengths)|Z) < 0.01 for generic related viruses.

(Wu 2023) counted the occurrence of (6,<8k).in a large collection of coronaviruses. Of 1316 betacoronaviruses, none except SC2 had (6,<8k), while 4 of 1378 more distantly related alphacoronaviruses did. Combining the two types, one estimates P((6,<8k)|Z) = 4/2694= 0.0015.

We should check that P((6,<8k)|Z) estimated from nearer relatives of SC2 is about the same as for the larger collections. Although (Wu 2023) found no instances in betacoronavuiruses, (Gadboit 2026) has tried generating clouds of mutant sequences around 5 known close relatives of SC2, with the number of random synonymous point mutations matched to their genetic distances from SC2, obtaining on average P((6,<8k)|Z) = 0.0042 but with large variations depending on the starting sequence. Extension to a larger collection of 17 relatives, some not as close, gave P((6,<8k)|Z) = 0.0057. (Gadboit 2026) Thus multiple estimates of P((6,<8k)|Z) based on actual existing viruses are all within a factor of two of 0.003, sufficiently accurate for our purposes here.

(Wu 2023) points out that over time SC2 variants drifted away from the initial pattern by picking up extra RE sites. That would be expected if the initial pattern was not an unlikely product of random evolution but rather an artifact of lab selection and manipulation. The loss of the pattern as a function of the number of mutations is close to that found in simulations of random synonymous mutations. (Gadboit 2026) The drift serves as a reminder that natural evolution is indifferent to the pattern.

The probability of obtaining the match of the pattern and the plan is higher under LL than under Z. Under Z, the only relevant causal path to SC2 is the one labeled "random" in Fig. 3, leaving roughly p=0.003. LL has several pathways, corresponding to techniques in (Almazán et al. 2014). The easiest paths would involve simple targeted point mutations to the potential sites. A reasonable estimate of P((6,<8k)|LL) might be roughly 0.1, which would leave a Bayes factor of over 20.

Under LL, the path labelled "choice of sequence" will have at least as high a probability as Z's "random". "Choice of sequence" will be enhanced by two effects. First, the plan could have been slightly adjusted (e.g. 6 vs. 7 segments) based on what the available sequences were at the time the grant was written. Second, the DEFUSE research group had more than 180 related sequences and would presumably, other things being about equal, choose the one that fit their RE site plan. If no suitable sequence was available the lab could have followed the "chimeric choice" plan which seems to be suggested by DEFUSE's discussion of making "consensus" sequences of fairly closely related viruses. Here again, other things being about equal, the choice could be somewhat guided by the planned RE site pattern. Unfortunately it is hard to subjectively estimate the probability of a lab using these methods.

(Hensel 2025) has argued that the sequences near key sites are all shared with other close relatives, consistent with the (6,<8k) arising by accident from recombination. To the uncertain extent that a lab choice at the level of a whole sequence of chimeric parts may have been influenced by the intended (6,<8k) pattern, the association with relatives' neighboring sequences would be the same for LL as for Z, but the more obvious way for LL to get (6,<8k) would be via synonymous point mutations. If one assumes that the chimeric structure of SC2 could not have been in part chosen by a lab to get the desired RE pattern, then this match with relatives needs to be treated as another observation, one which disfavors any LL paths involving targeted point mutations. Such mutations could break the association of the sequences at RE sites with the neighboring sequences of related viruses (Gadboit 2026) has now simulated the probability of chimeric structures being closely piecewise related to the known relatives, approximately estimating the probabilities based on how similar the SC2 neighborhood is to each relative in Hensel's set. When random synonymous point mutations reflecting the sequence differences are included, the probability of obtaining (6,<8kb) rises to between 11% and 13% under Z, using neighboring windows of from 50 to 100 nucleotides, following Hensel. That ~12% would be the same for any lab sequences. In other words, if the chimeric structure is assumed to have arisen purely under Z that accident already goes a long way toward giving the match between the plan and the RE pattern regardless of whether a lab handled the virus. Most of the 88% of the simulated sequences that do not give (6,<8k) fail by just one site, easy for a lab to fix by a synonymous mutation to get the intended pattern. If we assume just a 50% chance that a lab would choose to do that if given a starting sequence that was already close to the convenient RE pattern, we obtain a Bayes factor of
$BF= 1+ 0.88^2/(2*0.12) =4.2$.

The key point is that *under Z there is no non-random causal pathway leading from the collection of related viruses to the unusual RE pattern matching the DEFUSE plans*. The similarity would be a pure statistical accident. *Under LL there are several such causal pathways* that would increase the probability of a match between the plans and the observation although without access to lab records there is no clear way to choose which is most relevant. I think that it would be conservative to estimate
P(pattern match to plan|LL)/P(pattern match to plan|Z) = 4.2.
Although this factor is much less than would be obtained *prima facie* by comparing the natural rate of patterns closely matching the DEFUSE plans with the likelihood of following those plans it is best to be cautious with any factor involving new types of observation.

**7. THE HUANAN SEAFOOD MARKET**

The widespread impression that a zoonotic source has been demonstrated relies primarily on a high-profile pair of companion 2022 papers in Science (Jonathan E. Pekar et al. 2022; Worobey et al. 2022) that pointed to the Huanan Seafood Market (HSM) as the probable spillover site. A collection of informally published Bayesian analyses (Alexander 2024) that find that Z is more likely all depend on huge likelihood factors based on these HSM-related claims. (Jonathan E. Pekar et al. 2022) turns out to have erroneously reversed the conclusion implied by its data and model. With (Worobey et al. 2022) also running into major difficulties the sub-hypothesis $Z_{HSM}$ turns out to be an unlikely subset of the broader Z hypothesis. (Weissman 2026b) Further discussion of HSM-specific evidence is moved to an Appendix.

**8. OTHER FACTORS**

(X. Chen et al. 2024)discuss types of evidence that require some sociological judgment to evaluate. The most important is that WIV permanently shut down the data base from which outsiders could judge whether they had sequences that could give an SC2 chimera. UNC has successfully resisted FOIA requests for laboratory notes from Baric's lab. (Sartwell 2026) (Bloom 2021) discusses other ways in which data that could shed light on the origin have been hidden. A complete Bayesian risk assessment ought to include an LL-favoring factor based on these observations, but its estimation would be too subjective for use here. It can be left as an exercise for the reader.

**9. SUMMARY, WITH ROBUST ALLOWANCE FOR UNCERTAINTY**

At this point, our posterior odds favoring DEFUSE-related LL over Z are ~1700/1, summarized in Table 1. These include conservative allowances for uncertainty and omit some hard-to-estimate factors that might tend to favor LL.

Integrating over prior distribution of ln(prior odds) reduces the final odds from ~1700/1 to ~120/1. This remains a strong conclusion, one that I think is justified by the current data. This robust result is much less sensitive to changes in individual factors than are the raw odd.

Table 1 summarizes the rough Bayes results.

| **Feature** | **P(feature\|LL)** | **P(feature\|Z)** | **likelihood ratio** | **ln(ratio)** |
|---|---|---|---|---|
| **2019 Sarbeco priors** | **1/140** | **1/20** | **0.14** | **-2.0±2.3** |
| **Wuhan** | **0.8** | **1/100** | **80** | **4.4** |
| **no hosts** | **1** | **< 2/7** | **3.5** | **1.3** |
| **pre-adaptation** | **1/3** | **< 1/10** | **3.3** | **1.2** |
| **recent FCS insert** | **1/2** | **< 1/6** | **3** | **1.1** |
| **RE site pattern** | **see text** | **see text** | **4.2** | **1.4** |
| **Total** | | | **1700** | **7.4 ± 2.3** |
| **Robust: 3-d.o.f. t** | | | **110** | |

This result agrees qualitatively with the mGFT analysis (X. Chen et al. 2024) and with ((Bannan 2026) but now includes rough odds. It differs quantitatively from the much stronger result obtained by (Quay 2026) because I included likelihood factors for LL rather than just for Z, did not limit the Z hypothesis to its particularly unlikely $Z_{HSM}$ subset, used more caution with potentially large factors, avoided double-counting of correlated factors, omitted factors that required much sociological judgment, and used robust averaging over uncertain priors. It differs from some analyses that favor Z (Alexander 2024) for several reasons, of which the most important is that those analyses include a huge Z-favoring factor based on the existence of the HSM case cluster, without which they would favor LL. Thus those Z-favoring Bayesian results are entirely reliant on the $Z_{HSM}$ sub-hypothesis, using an abstract model to calculate odds of a near-market case cluster rather than using empirical comparisons and allowing for known limitations of the data. They also fail to take into account how small the Wuhan share of the wildlife trade is or some other ways in which $Z_{HSM}$ does not fit the data. (Weissman 2026b) They do not include the close match of the RE site pattern to DEFUSE plans.

**11. CONCLUSION**

Based on the currently available unclassified data even a conservative estimate of the odds strongly favors a research-related accidental leak over a zoonotic source. As with any probabilistic analysis, the current results should not be taken as the final word. New evidence may come in.

Nevertheless, at this point the results point toward the need for better regulation of experiments on modifying potential pandemic pathogens. (A. Chan 2026) The general direction of regulation needed to avoid the greatest risks while minimizing restrictions of useful research was outlined before the pandemic. (Lipsitch 2018; Lipsitch and Inglesby 2014) Since potential pandemic pathogen research occurs in numerous countries and pandemic pathogens do not respect political boundaries, effective regulation will require an international scope.

**APPENDIX ON THE HUANAN SEAFOOD MARKET EVIDENCE**.

Recent papers (Bloom 2025, 2023; Weissman 2026a, 2026b) have described serious problems with claims that evidence points to $Z_{HSM}$. In this Appendix I briefly summarize the evidence that $Z_{HSM}$ is substantially less likely than a more generic Z, regardless of the prior probability assigned to it. I will assign very rough Bayes factors to the evidence, but that should suffice here.

$Z_{HSM}$ starts with a much reduced $P(\text{Wuhan}|Z_{HSM})$ compared to generic $P(\text{Wuhan}|Z)$. The wet-

market human animal interface in Wuhan is a particularly low fraction of the national interface. For example, the ~500 bamboo rats sold each year in Wuhan markets (Xiao et al. 2021) account for a tiny fraction of the national production. When the pandemic hit Guangxi province alone had 36,000,000 bamboo rats, most of which were to be harvested at age less than 10 months. (Lin and Rupeng 2020) Similar numbers for raccoon dogs and other animals that have been mentioned as possible hosts (Xiao et al. 2021; Chinese_Academy_of_Engineering 2017) indicate that, conservatively, $P(Wuhan|Z_{HSM}) < 0.001$. Thus evidence for $Z_{HSM}$ would need to be strong for it to boost P(Z) overall. Roughly speaking, $P(Wuhan|Z_{HSM})/ P(Wuhan|Z) < 0.1$

One of the key Science companion papers (Jonathan E. Pekar et al. 2022) supporting $Z_{HSM}$ concluded via a series of errors in coding and logic (McCowan 2025) that there were probably two successfully propagating spillovers rather than one. Minimal repair of those errors (Weissman 2026a) using only the data, model, and simulations of the original paper reverses the conclusion. Correcting this error affects some of the claims regarding $Z_{HSM}$ but does not lead directly to a substantial Bayes factor.

The other Science paper (Worobey et al. 2022) points out that the Dec. 2019 superspreading event at the Huanan Seafood Market would be more likely if the virus originated there rather than elsewhere. Nevertheless, superspreading events at wet markets are not especially rare, e.g. the spread from Xinfadi market after 56 days without local spread in Beijing (Dai et al. 2023), or seafood markets in Singapore (Leo 2021) and a suburb of Bangkok (Harmer 2020). Since there are only about 40 cities in Asia as large as Singapore and 3 had wet market outbreaks after quiet periods the mere presence of a superspreading event at a market is only about 15 times more likely if the market is the actual spillover source than it is for markets that pick up the virus from human visitors. Thus this market-linked outbreak favors a market origin:

$P(HSM|Z_{HSM})/ P(HSM|Z) = \sim 15$

With that likelihood factor approximately cancelling the $P(Wuhan|Z_{HSM})/P(Wuhan|Z)$ factor one needs to consider other factors to see if the $Z_{HSM}$ is overall favored or disfavored compared to generic Z.

1. Multiple independent lines of evidence indicate that the first cases occurred well before any were found at HSM. (Wenjun Wang et al. 2020; Rainey and Feldman 2020; Borrell 2020; Fouchier 2020; Ma 2021; Birrell 2021; J. Pekar et al. 2021; Roberts, Rossman, and Jarić 2021; La Rosa et al. 2021; Washington_Post_Editorial_Board 2022; Caraballo-Ortiz et al. 2022; Demaneuf 2024; Committee on Oversight and Responsibility 2024)

Under $Z_{HSM}$ it seems highly unlikely that a series of independent Western experts would all report hearing of the new outbreak before that rumor could possibly have arisen from the reported HSM cases. The other types of evidence cited, especially the large discrepancy betwen the early cases in early Chines papers an the later reports of cases that included ones at HSM, make that even more unlikely. A conservative estimate would be
$P(\text{early reports}|Z_{HSM})/ P(\text{early reports}|Z) < 0.1$

2. Unlike the known Beijing market event where almost all the cases were linked to the Xinfadi market (Dai et al. 2023), many of the early official Wuhan cases were not linked to HSM. (WHO-China 2021) This contrast seems highly anomalous if the there were not Wuhan cases originating outside the market.
$P(\text{early unlinked cases}|Z_{HSM})/ P(\text{early unlinked cases}|Z) < 0.1$

3. All the HSM cases whose lineage was determined were Lineage B, farther from the natural

relatives than Lineage A and several variants of A that were spreading at other locations. (Bloom 2025). That is reversed from what would be expected under $Z_{HSM}$. Attempts to explain the reversal by positing multiple spillovers(Jonathan E. Pekar et al. 2022) were erroneous. (Weissman 2026a)

P(non-ancestral lineage|$Z_{HSM}$)/ P(non-ancestral lineage cases|Z) = ~0.5

4. No animal at HSM tested positive for SC2 out of 457 sampled. (W.J. Liu et al. 2024)

That is clearly unexpected if the prior host was in the market.

P(no HSM host found|$Z_{HSM}$)/ P(no HSM host found |Z) = ~0.5

5. SC2 RNA on environmental swabs taken in HSM (W.J. Liu et al. 2024) was not positively correlated with mtDNA from suspected hosts, in contrast to 4 other prevalent coronaviruses whose RNA did correlate positively with their hosts' mtDNA. (One other RNA virus showed no positive correlation with any likely host but it was present at especially low levels.) The SC2 RNA correlated strongly with human mtDNA. (Bloom 2023)

This strengthens the case that no host was present in HSM.

P(no mtDNA-RNA association|$Z_{HSM}$)/ P(no mtDNA-RNA association|Z) = ~0.5

6. No SC2-related viruses other than SC2 were found in HSM swabs, indicating that live animals from the suspected zoonotic source region were not likely to be sold there, as also indicated by direct source records. (W.J. Liu et al. 2024; WHO-China 2021; Bloom 2023) This strengthens the previous points but may overlap too much to justify another Bayes factor.

The product of our crude Bayes factors favors Z over $Z_{HSM}$ by about a factor of 500. Thus overall HSM is particularly unlikely to be the spillover site. That only slightly reduces the posterior probability of Z since not too much of the prior probability of Z needs to have been assigned to

$Z_{HSM}$.